\documentclass{article}
\usepackage{amsfonts,amssymb, amsmath}
\usepackage[english]{babel}

\DeclareMathAlphabet{\mathpzc}{OT1}{pzc}{m}{it}

\def\nn{\nonumber }
\def\bq{ \begin{equation} }
\def\eq{ \end{equation} }
\def\ben{ \begin{eqnarray} }
\def\en{ \end{eqnarray} }

\def\ii{{\rm i}}
\def\ee{{\rm e}}

\newtheorem{re}{Remark}
\newenvironment{rem}{\begin{re} \rm }{\end{re}}
\begin{document}


\title{New variables of separation for  particular case of the Kowalevski top.}
\author{ A V Tsiganov \\
\it\small
St.Petersburg State University, St.Petersburg, Russia\\
\it\small e--mail:  andrey.tsiganov@gmail.com}

\date{}
\maketitle

\begin{abstract}
We  discuss the polynomial bi-Hamiltonian structures for the Kowalevski top
in  special case of zero square integral.
An  explicit procedure to find  variables of separation
and separation relations is considered in detail.
\end{abstract}


\vskip0.1truecm
\begin{flushright}
To V.V. Kozlov on the occasion of his 60th birthday
\end{flushright}
\vskip0.1truecm

\section{Introduction}
\setcounter{equation}{0}
During a century the only cases of integrability of the Euler-Poisson equations were the isotropic case and the cases of Euler (1758) and Lagrange (1788). In 1888
S. Kowalevski  found a new  highly non-trivial case of integrability \cite{kow89}. In modern terms, this
is an integrable system on the e(3) algebra with  quadratic and  quartic
(in angular momenta) integrals of motion.

Furthermore, by using a mysterious change of variables, she showed that  equations of motion for the new case of integrability are linearized on the abelian variety by means of the Jacobi-Abel theorem about the inversion of a system of abelian integrals \cite{kow89}. At the moment no separation which is alternative to her original separation of variables is known for this system
, even though there is a large body of literature dedicated to the
problem, including the detailed geometric description
of the invariant surfaces on which the motion evolves, see books
\cite{au96,bm05} and references within.

In this paper we discuss the direct method of finding variables of separation without any additional information (ingenious and at times obscure change of variables, Lax matrices, $r$-matrices, links with soliton equations  etc). For example,  we apply the machinery of bi-Hamiltonian geometry to  the Kowalevski top at  zero level of the cyclic integral of motion, which is a particular case of the generic Kowalevski top. The rational Poisson bivector associated with  famous Kowalevski variables may be found in \cite{ts08c}.
Here we specially do not consider Kowalevski variables in order to get  only the new variables of separation and  the new underlying polynomial Poisson structures.

The other aim is  the construction of different variables of separation  lying on the distinct  algebraic curves \cite{ts10}. Relations between such distinct curves  give us a lot of new examples of reductions of Abelian integrals and, therefore, they may be the source  of new ideas in the number theory, algebraic geometry and modern cryptography \cite{au96,curv3,mark01}.

In Section 2 we construct new compatible Poisson bivectors for the Kowalevski top. In Section 3 we find the new corresponding
variables of  separation and the separated relations.
Finally, some concluding remarks can be found in the last Section.

\section{The bi-hamiltonian structure}
\setcounter{equation}{0}

\subsection{Description of the model}
According to \cite{kow89}, the Kowalevski top is  a dynamical system
with the following  integrals of motion \ben\label{kow-H}
H_1&=& J_1^2+J_2^2+2J_3^2+c_1x_1,\qquad\qquad\qquad c_1\in\mathbb R,\\
\nn\\
H_2&=&(J_1^2+J_2^2)^2-2\Bigl(x_1(J_1^2-J_2^2)+2x_2J_1J_2\Bigr)c_1+(x_1^2+x_2^2)c_1^2.\nn
\en
Here $J_i$ are the components of the angular momentum in  the moving frame of coordinates attached to the principal axes of inertia. The position of a rigid body is fixed by the components $x_i$ of the Poisson vector, which are the cosines between the axes of the body frame and the field up to a constant.

Using the Hamilton function $H_1$ and the Lie-Poisson bracket $\{.,.\}$
on  the Euclidean algebra $e^*(3)$ the customary Euler-Poisson  equations may be rewritten
in the hamiltonian form
\bq\label{eq-mkov}
\dot{J}_i=\{J_i,H_1\}\,,\qquad \dot{x}_i=\{x_i,H_1\}\,,\qquad\mbox{where}\qquad\{f,g\}=\langle Pdf,dg\rangle\,.
\eq
In coordinates  $z=(x_1,x_2,x_3,J_1,J_2,J_3)$ on
 $e^*(3)$   the Lie-Poisson bivector $P$  is the following antisymmetric matrix
\[
P=\left(
  \begin{array}{cccccc}
    0 & 0 & 0 & 0 & x_3 & -x_2 \\
    * & 0 & 0 & -x_3 & 0 & x_1 \\
    * & * & 0 & x_2 & -x_1 & 0 \\
    * & * & * & 0 & J_3 & -J_2 \\
    * & * & * & * & 0 & J_1 \\
    * & * & * & * & * & 0 \\
  \end{array}
\right).
\]
It has  two Casimir elements
\bq \label{caz-e3}
PdC_{1,2}=0,\qquad C_1=|x|^2\equiv\sum_{k=1}^3 x_k^2, \qquad C_2= \langle x,J \rangle\equiv\sum_{k=1}^3 x_kJ_k.
\eq
After fixing values of the Casimir elements
\[C_1=a^2,\qquad C_2=b\]
one gets a generic four-dimensional  symplectic leaf $\mathcal O_{ab}$, which is topologically equivalent to the cotangent bundle $T^*\mathcal S^2$ of the sphere $\mathcal S^2$ with radius $a$. However, the symplectic structure of $\mathcal O_{ab}$ is different from the standard symplectic structure on $T^*\mathcal S^2$ by the magnetic term proportional to $b$ \cite{nov81}.

The Kowalevski top is an integrable system on the phase space $\mathcal O_{ab}$ because the two independent integrals of motion $H_{1,2}$ (\ref{kow-H}) are in the involution
\bq\label{inv-H}
\{H_1,H_2\}=\langle PdH_1,dH_2\rangle=0.
\eq
In mechanics the Casimir function $C_1$ is a norm of the unit Poisson vector such as $a=1$,
whereas  second Casimir function $C_2$ is called a square  or cyclic integral of motion \cite{bm05,yeh}.

\begin{rem}
In original Kowalevski work the first step in the separation of variables method
 is the complexification: she considers
\[
\mathrm z_1=J_1 + i J_2, \qquad \mathrm z_2=J_1-i J_2
\]
as independent complex variables. Next she makes
her famous change of variables
\[
\label{s1,2} s_{1,2}=\frac{R(\mathrm z_{1} ,\mathrm z_{2} ) \pm\sqrt{R(\mathrm z_{1} ,\mathrm z_{1}
)R(\mathrm z_{2} ,\mathrm z_{2} )}} {2(\mathrm z_{1} -\mathrm z_{2} )^2}.
\]
The fourth degree polynomials $R(\mathrm z_i,\mathrm z_k)$ we will not specify here.
It brings the system (\ref{eq-mkov}) to the form
\[
(-1)^k\,(s_1-s_2)\dot{s}_k=\sqrt{P(s_k)\,}\,,\qquad
k=1,2,
\]
where
\bq\label{Kow-pol}
P(s)=4\left((s-H)^2-\frac{K}{4}\right)\left[s\left( (s-H)^2+c_1^2a^2-\dfrac{K}{4}\right)+c_1^2b
\right]
\eq
Here the pairs $\left(s_k, \eta_k=\sqrt{P(s_k)}\right)$,
can be regarded as coordinates of points on the Kowalevski curve of genus two
\bq
\mathcal C_{kow}:\qquad\eta^2-P(s)=0\,.
\eq
At  zero level of the cyclic integral of motion $C_2=0$ the Kowalevski curve has the same genus $2$.
\end{rem}

We address the problem of separation of variables for the Hamilton-Jacobi equation as well.
At $C_2=0$ the symplectic leaves $\mathcal O_{a0}$  are completely symplectomorphic to  $T^*{\mathcal S}^2$ \cite{nov81}.  We will only consider such symplectic leaves and, therefore, all the formulae below hold true up to $C_2=0$.

For this Kowalevski top  on the two-dimensional
sphere we want to calculate different variables of separation and,
according to  the general usage of the bi-hamiltonian geometry, firstly we have to find the second dynamical Poisson bivector $P'$ equipped with some necessary properties \cite{fp02,mag97}.

\subsection{Dynamical Poisson bivectors}

According to  \cite{ts07c,ts08,ts08b,ts09} let us suppose that the desired second Poisson bivector $P'$ is the Lie derivative of $P$ along some unknown Liouville vector field $X$
\bq\label{co-b}
P'=\mathcal L_X(P)\,.
\eq
In addition it has to satisfy  the following equations
\bq\label{m-eq1}
[P',P']\equiv[\mathcal L_X(P),\mathcal L_X(P)]=0,
\eq
and
\bq\label{m-eq2}
\{H_1,H_2\}'=\langle P'dH_1,dH_2\rangle=0,
\eq
where $[.,.]$ is the  Schouten bracket.

The first assumption (\ref{co-b}) guarantees that this dynamical  bivector $P'$ is compatible with the given kinematic Poisson bivector $P$, i.e. $[P,P']=0$. In geometry such bivector $P'$ is said to be the 2-coboundary associated with the Liouville vector field  $X$ in the Poisson-Lichnerowicz cohomology defined by $P$.

The second condition (\ref{m-eq1}) means that $P'$ is the Poisson bivector, i.e. that the Jacobi identity is true. The third equation (\ref{m-eq2}) relates $P'$ with the given integrable system. In the wake of this agreement the foliation defined by the $H_{1,2}$ is the bi-Lagrangian foliation \cite{fp02,mag97}.

The system of equations (\ref{m-eq1}-\ref{m-eq2}) has infinitely many solutions with respect to $X$ \cite{ts07a,ts08b}. So, in order to get some particular solution we have to narrow the search space. In this paper we suppose that
\bq\label{m-eq3}
P'dC_{1,2}=0,
\eq
and that the components $X_j$ of the Liouville vector field $X=\sum X_j\,\partial_j$ are non-homogeneous polynomials in momenta $J_k$
\[
X_j=\sum_{m=0}^N\sum_{k=0}^m \mathrm g_{jkm}^{N}(x_1,x_2,x_3)\,J_1^k\,J_2^{m-k}
\]
with unknown coefficients $\mathrm g(x_1,x_2,x_3)$ \cite{ts07c,ts08,ts09}. Here we explicitly use the restriction $C_2=0$, i.e. that $J_3=-(x_1J_1+x_2J_2)/x_3$.

Upon substituting this polynomial \textit{ans\"{a}tze} into the equations (\ref{m-eq1},\ref{m-eq2}-\ref{m-eq3}) and demanding that all the coefficients at powers of $J_k$ vanish one gets the over determined system of  algebro-differential equations. Such systems are  solved on  personal computer by using modern software in a few seconds. So, the only real problem is the classification and the analysis of the received computer results.

The first three nontrivial solutions  arise only in the cubic case $N=3$. Components of the first real vector field $X^{(1)}$ are equal to
\ben
X^{(1)}_1&=&-\frac{\sqrt{x_1^2+x_2^2}\Bigl(x_1J_1-x_2J_2\Bigr)J_3}{2x_1x_3},\qquad X^{(1)}_2=\frac{\sqrt{x_1^2+x_2^2}\Bigl(x_1J_1-x_2J_2\Bigr)J_3}{2x_2x_3},\qquad X^{(1)}_3=0,\nn\\
\label{kow-sol1}
X^{(1)}_4&=&-\sqrt{x_1^2+x_2^2}\left(\frac{(x_1^2+x_2^2)J_1^3}{6x_2^2x_3^2}-\frac{J_2^2J_3}{2x_1x_3}\right)+
\frac{c_1x_3J_3}{4\sqrt{x_1^2+x_2^2}}\,,\\
X^{(1)}_5&=&-\sqrt{x_1^2+x_2^2}\left(\frac{(x_1^2+x_2^2)J_2^3}{6x_1^2x_3^2}-\frac{J_1^2J_3}{2x_2x_3}   \right)+\frac{c_1(x_1J_2-x_2J_1)}{4\sqrt{x_1^2+x_2^2}}\,,\nn\\
X^{(1)}_6&=&-\sqrt{x_1^2+x_2^2}\frac{(x_1^2+x_2^2)J_3^3}{6x_1^2x_2^2}+\frac{c_1\sqrt{x_1^2+x_2^2}J_1}{4x_3}\,.\nn
\en
The components of the second real vector field $X^{(2)}$ read as
\ben
X^{(2)}_1&=&\frac{2(x_1^2+x_2^2)}{x_3}\,J_1J_3\,,\qquad\qquad X^{(2)}_2=-\frac{2x_1(x_1^2+x_2^2)}{x_2x_3}\,J_1J_3\,,\qquad \qquad X^{(2)}_3=0\,,\nn\\
X^{(2)}_4&=&\frac{(x_1^2+x_2^2)^2}{3x_2^2x_3^2}\,J_1^3-\left(J_1+\frac{x_1x_3}{3x_2^2}J_3\right)J_3^2+\frac{x_1^2+x_2^2}{x_2x_3}\,J_1J_2J_3
+\frac{c_1x_2J_2}{2}\,,
\nn\\
\label{kow-sol2}
X^{(2)}_5&=&
\dfrac{(x_1^2+x_2^2)^2}{3x_1^2x_3^2}\,J_2^3
-\left(J_2-\frac{(2x_1^2-x_2^2)x_3}{3x_1^2x_2}\,J_3-\frac{2(x_1^2+x_2^2)}{x_1x_2}J_1\right)J_3^2
\\
&+&\frac{x_1^2+x_2^2}{x_1x_3}\,J_1J_2J_3-\frac{c_1(2x_1J_2-x_2J_1)}{2}\nn\\
\nn\\
X^{(2)}_6&=&\frac{2x_1^2+x_2^2}{3x_2^2}\,J_3^3+\frac{c_1x_2(x_1J_2-x_2J_1)}{2x_3}\nn
\en
The components of the third vector field $ X^{(3)}$ are the complex functions  on initial variables
\ben
 X^{(3)}_1&=&-\frac{\ii x_2(x_1+\ii x_2)^2}{x_1^2}\,J_2^2+\frac{2x_2(x_1+\ii x_2)}{x_1}\,J_1J_2,\qquad\qquad \ii=\sqrt{-1}\,,\nn\\
\label{kow-sol3}
 X^{(3)}_2&=&\frac{\ii(x_1+\ii x_2)^2}{x_1}J_2^2-2(x_1+\ii x_2)J_1J_2,\qquad X^{(3)}_3=0\,,\\
 X^{(3)}_4&=&
(J_1-\ii J_2)J_2^2-\frac{1}{3}J_1^3+\frac{2(2x_1+\ii x_2)x_3}{3x_1^2}\,J_3^3+\frac{4x_2}{3x_1}J_2^3+
\frac{\ii x_2 (x_1J_1-x_3J_3)}{x_1^2}J_2^2\nn\\
&+&c_1x_3J_3\,,
\nn\\
 X^{(3)}_5&=&\frac{2\ii}{3}J_1^3-(J_1-\ii J_2)J_1J_2-\frac{1}{3}J_2^3-\frac{2\ii x_3}{3x_1}J_3^3+\frac{x_2^2(2x_1-\ii x_2)}{3x_1^3}J_2^3-\frac{\ii x_2^2x_3}{x1^3}J_2^2J_3
\nn\\
&+&\ii c_1x_3J_3\,,
\nn\\
 X^{(3)}_6&=&\frac{2}{3}\frac{(x_1+\ii x_2)x_3^2-2x_1^3}{x_1^3}\,J_3^3-(J_1^2+J_2^2)\,J_3
-c_1(x_1+\ii x_2)\,J_3\,.\nn
\en
The quartic ans\"{a}tze  yields a lot of solutions, which will be classified and
studied in future.

Let us show the simplest part of these real and complex Poisson brackets explicitly
\[\begin{array}{ll}
\{x_i,x_j\}=\varepsilon_{ijk}\,x_k\,,\qquad &
\{x_i,x_j\}^{(1)}=\varepsilon_{ijk}\dfrac{\sqrt{x_1^2+x_2^2}\,J_3}{x_3}\,x_k\,,\\
\{x_i,x_j\}^{(2)}=-\varepsilon_{ijk}\dfrac{2(x_1^2+x_2^2)\,J_3}{x_3}\,x_k\,,\qquad&
\{x_i,x_j\}^{(3)}=2\ii\varepsilon_{ijk}\,(\ii x_3J_3-x_1J_2+x_2J_1)x_k\,.
\end{array}
\]
Here $\varepsilon_{ijk}$ is the totally skew-symmetric tensor. Other brackets are appreciably more tedious expressions. The complex Poisson structure may be rewritten in the lucid form by using the $2\times 2$ Lax matrices  \cite{kuzts89,ts02} and  the bi-hamiltonian structure associated with the reflection equation algebra \cite{ts08c}.

It is easy to prove that the corresponding Poisson bivectors $P^{(1)}$, $P^{(2)}$ and $P^{(3)}$ have the following properties
\bq\label{comp-P12}
[P^{(1)},P^{(2)}]=0,\qquad [P^{(1)},P^{(3)}]\neq 0,\qquad [P^{(2)},P^{(3)}]\neq 0\,
\eq
with respect to the Schouten brackets. It means that $P^{(1)}$ and $P^{(2)}$ are compatible bivectors, whereas the complex bivector  ${P}^{(3)}$ is incompatible with the real bivectors.

\begin{rem}
For any bivectors $P$ and  $P'$ there are a lot of vector fields $X$, such as  $P'=\mathcal L_X(P)$. Above we put $X_3=0$ in order to restrict this freedom. It may be the origin of some non-symmetry and irregularity in expressions (\ref{kow-sol1},\ref{kow-sol2}) and (\ref{kow-sol3}).
\end{rem}

\begin{rem}
There are two \textit{rational} Poisson bivectors $P'$ for the Kowalevski top. The first bivector is associated with the Kowalevski variables of separation and the underlying reflection equation algebra \cite{ts08c}. The second bivector is related with the Lax matrix of Reyman-Semenov-Tian-Shansky and the linear $r$-matrix algebra \cite{ts07}. The components of the corresponding vector fields $X$ are logarithmic functions in momenta.
\end{rem}

To sum up, using the applicable polynomial  ans\"{a}tze for the
Liouville vector field $X$  we got two compatible real
cubic bivectors $P^{(1,2)}=\mathcal L_{X^{(1,2)}}(P)$  and one complex
cubic bivector $P^{(3)}=\mathcal L_{X^{(3)}}(P)$ for the Kowalevski top.
Although these bivectors are defined by arbitrary value of $C_2$,
they are compatible with the  initial Poisson bivector $P$ only for $C_2=0$.
The  application of this Poisson bivectors will be given in the next section.

\section{Calculation of the variables of separation and the separation relations}
\setcounter{equation}{0}
A system of canonical variables $(q_1,\dots,q_n,p_1,\dots,p_n)$
\bq
 \{q_i,q_k\}=\{p_i,p_k\}=0,\qquad  \{q_i,p_k\}=\delta_{ik}
\eq
will be called {\it separated} if there are $n$ relations of the form
\begin{equation}
\label{seprelint}
\Phi_i(q_i,p_i,H_1,\dots,H_n)=0\ ,\quad i=1,\dots,n\ ,
\qquad\mbox{with }\det\left[\frac{\partial \Phi_i}{\partial H_j}\right]
\not=0\,,
\end{equation}
binding together each pair $(q_i, p_i)$ and  $H_1,\ldots,H_n$.

The reason for this definition is that the stationary
Hamilton-Jacobi equations for the Hamiltonians $H_i=\alpha_i$ can be
collectively solved by the additively separated complete integral
\begin{equation}\label{eq:i0}
W(q_1,\dots,q_n;\alpha_1,\dots,\alpha_n)=
\sum_{i=1}^n W_i(q_i;\alpha_1,\dots,\alpha_n)\>,
\end{equation}
where  $W_i$ are found by the quadratures as the solutions of the ordinary
differential equations.

The integrals of motion $(H_1,\dots,H_n)$ are the St\"ackel  separable integrals
if the separation relations (\ref{seprelint}) are given by affine equations in  $H_j$, that is,
\begin{equation}
\label{stseprel}
\sum_{j=1}^n S_{ij}(q_i,p_i) H_j-U_i(q_i,p_i)=0\ ,\qquad
i=1,\dots,n\ ,
\end{equation}
with  an invertible matrix $S$.  The functions $S_{ij}$ and
$U_i$ depend only on one pair  $(q_i,p_i)$ of the canonical variables of separation, it means that
\bq\label{st-mat}
\{S_{ik}, q_j\}=\{S_{ik}, p_j\}=\{ S_{ik},S_{jm} \}=0,\qquad i\neq j,
\eq
and similar to $U$
\bq\label{st-pot}
\{U_{i}, q_j\}=\{U_{i}, p_j\}=\{ U_{i},U_{j} \}=0,\qquad i\neq j.
\eq
In this case $S$ is called the {\em St\"ackel matrix\/}, and $U$ the {\em St\"ackel potential\/}.

\begin{rem}
We have to point out that the definition of the St\"ackel separability depends on the choice of  $H_i$. Indeed, if $(H_1,\dots,H_n)$ are St\"ackel-separable, then
$\widehat{H}_i=\widehat{H}_i(H_1,\dots,H_n)$ will not, in general,
fulfill the affine relations of the form (\ref{stseprel}).
\end{rem}

\begin{rem}
The method of the separation of variables for a long time
served an important, but technical role in solving Liouville
integrable systems of classical mechanics.
A new, and much more exciting application of the method came with
the development of quantum integrable systems. Because of the fact
that the quantization of the
action variables seems to be a rather formidable task, quantum separation
of variables became an inevitable refuge. In fact, it can be successfully
performed for many families of integrable systems with affine separated relations (\ref{stseprel})
 only \cite{Skl,sm00}.
\end{rem}

So, our second step  is the calculation of the canonical variables of separation $(q_i,p_i)$ and of the separation relations $\Phi_i$ (\ref{seprelint}).
According to \cite{fp02,mag97}, the coordinates of separation $q_i$ are eigenvalues of the recursion operator, which are the so called  Darboux-Nijenhuis variables. In order to get the recursion operator $N=\widehat{P}'\widehat{P}^{-1}$ we have to find restrictions $\widehat{P},\widehat{P}'$ of the Poisson bivectors $P$ and $P'$ onto the symplectic leaves.

We can avoid the procedure of restriction using the $n\times n$ control matrix $F$ defined by
\bq\label{f-mat}
P'{\mathbf{dH}}=P\bigl(F{\mathbf{dH}}\bigr),\qquad\mbox{or}\qquad
P'dH_i=P\sum_{j=1}^n F_{ij}\,dH_j,\qquad i=1,\ldots,n.
\eq
The bi-involutivity of the integrals of motion (\ref{inv-H},\ref{m-eq2})
is equivalent to the existence of $F$, whereas the imposed condition (\ref{m-eq3}) ensures that $F$ is a non-degenerate matrix. In this case  eigenvalues of this matrix coincide with the Darboux-Nijenhuis variables and we can easily {\it calculate} the desired coordinates of separation $q_i$.

Moreover, for the St\"ackel separable systems the suitable normalized left eigenvectors of the control matrix $F$ form the St\"ackel matrix $S$
 \[
 F=S^{-1}\,\mbox{diag}\,(q_1,\ldots,q_n)\,S
 \]
which would allow us to get separated relations (\ref{stseprel}).

So, the main problems are the finding of the conjugated momenta $p_{i}$ and the construction of the separation relations $\phi_j$ (\ref{seprelint}) for the generic non-St\"ackel separable systems. Below we show how we can solve these problems using the same control matrix $F$ and some additional useful observations.

\subsection{The real compatible Poisson bivectors}

For the first Poisson bivector $P^{(1)}$ (\ref{kow-sol1}) the entries of the control matrix $F^{(1)}$ read as
\ben
F^{(1)}_{11}&=& \frac{(2x_1^2+2x_2^2+x_3^2)(J_1^2+J_2^2)}{4x_3^2\sqrt{x_1^2+x_2^2}}\qquad F^{(1)}_{12}=-\frac{1}{8\sqrt{x_1^2+x_2^2}} \nn\\
F^{(1)}_{21}&=&
   \frac{(2x_1^2+2x_2^2+x_3^2)(J_1^2+J_2^2)}{2x_3^2\sqrt{x_1^2+x_2^2}}
-\frac{c_1\sqrt{x_1^2+x_2^2}\bigl(x_1(J_1^2-J_2^2)+2x_2J_1J_2\bigr)}{x_3^2}
-\frac{c_1^2\sqrt{x_1^2+x_2^2}}{2}\nn\\
F^{(1)}_{22}&=& -\frac{J_1^2+J_2^2}{2\sqrt{x_1^2+x_2^2}} \nn
\en
The eigenvalues $q_{1,2}$ of this matrix are the required variables of separation $q_{1,2}$
\ben\label{q1-def}
\det(F^{(1)}-\lambda I)&=&(\lambda-q_1)(\lambda-q_2)\nn\\&=&\lambda^2
-\dfrac{\sqrt{x_1^2+x_2^2}(J_1^2+J_2^2)}{2x_3^2}\lambda
-\dfrac{c_1\Bigl(2x_1(J_1^2-J_2^2)+4x_2J_1J_2+c_1x_3^2\Bigr)}{16x_3^2}
.\nn
\en
The matrix of normalized eigenvectors of $F^{(1)}$ does not form the St\"ackel matrix, because property (\ref{st-mat}) is missed, and  the underlying separation relations differ from the St\"ackel affine equations (\ref{stseprel}) in $H_{1,2}$.

For the second Poisson bivector $P^{(2)}$ (\ref{kow-sol2}) the entries of the control matrix $F^{(2)}$ are equal to
\ben
F^{(2)}_{11}&=&-\dfrac{J_1^2+J_2^2-c_1x_1}{2}+\frac{(x_1J_2-x_2J_1)^2}{x_3^2}\,,\qquad  F^{(2)}_{12}=\frac{1}{4}\,,
\nn\\
F^{(2)}_{21}&=&-(J_1^2+J_2^2)^2\left(1+\frac{2(x_1^2+x_2^2)}{x_3^2}\right)
+c_1(x_1^2+x_2^2)\left(
\frac{2\Bigl(x_1(J_1^2-J_2^2)+2x_2J_1J_2\Bigr)}{x_3^2}
+c_1\right)\,,
\nn\\
F^{(2)}_{22}&=& \dfrac{J_1^2+J_2^2+2J_3^2+c_1x_1}{2}\,.\nn
\en
The eigenvalues $f_{1,2}$ of the matrix $F^{(2)}$ are the roots of the equation
\ben\label{q2-def}
\det(F^{(2)}-\lambda I)&=&(\lambda-f_1)(\lambda-f_2)\nn\\&=&\lambda^2
-\left(c_1x_1-\frac{(x_2J_1-x_1J_2-x_3J_3)(x_2J_1-x_1J_2+x_3J_3)}{x_3^2}\right)\lambda\nn\\
&-&\frac{(2(x_2J_1-x_1J_2)J_3-c_1x_2x_3)^2}{4x_3^2}\,.\nn
\en

\begin{rem}
According to \cite{fp02,mag97} the compatibility of ${P}^{(1,2)}$ (\ref{comp-P12}) ensures that in the Darboux-Nijenhuis variables $q,p$ the corresponding restrictions of $\widehat{P}^{(1,2)}$ look like
\[
\widehat{P}^{(1)}=\left(
  \begin{array}{cccc}
    0 & 0 & q_1 & 0 \\
    0 & 0 & 0 & q_2 \\
    -q_1 & 0 & 0 & 0 \\
    0 & -q_2 & 0 & 0
  \end{array}
\right),\qquad
\widehat{P}^{(2)}=\left(
  \begin{array}{cccc}
    0 & 0 & f_1 & 0 \\
    0 & 0 & 0 & f_2 \\
    -f_1 & 0 & 0 & 0 \\
    0 & -f_2 & 0 & 0
  \end{array}
\right).
\]
where $f_{1,2}$ are the functions on $q,p$ such as
\[ \{q_i,f_j\}=\{p_i,f_j\}=0,\qquad i\neq j\,.\]
So, $f_1$ is the function only on $q_1$ and $p_1$ and similar  $f_2$ is the function on $q_2$ and $p_2$.

\end{rem}

We can find these functions $f_{1,2}$ using the Poisson bracket.
Namely, it is easy to see that the recurrence chain
\bq\label{rrel1}
\phi_1=\{f_1(q_1,p_1),q_1\},\quad \phi_2=\{\phi_1,q_1\},\quad
\ldots,\quad \phi_i=\{\phi_{i-1},q_1\}
\eq
breaks down on the third step $\phi_3=0$. It means that ${f}_1$ is the second order polynomial in momenta $p_1$ and, therefore, we can define this unknown momenta in the following way
\bq\label{pkow-def1}
p_1=\dfrac{\phi_1}{\phi_2}=
\dfrac{2x_3\Bigl(4(x_2J_1-x_1J_2)\,q_1+c_1\sqrt{x_1^2+x_2^2}J_2\Bigr)}
{(4\sqrt{x_1^2+x_2^2}(J_1^2+J_2^2)\,q_1+c_1\Bigl(x_1(J_1^2-J_2^2)+2x_2J_1J_2\Bigr)}
\eq
up to  canonical transformations $p_1\to p_1+\mathrm g(q_1)$.

The similar calculation for the function $f_2(q_2,p_2)$ yields the definition of the second momenta
\bq\label{pkow-def2}
p_2=\dfrac{2x_3\Bigl(4(x_2J_1-x_1J_2)\,q_2+c_1\sqrt{x_1^2+x_2^2}J_2\Bigr)}
{(4\sqrt{x_1^2+x_2^2}(J_1^2+J_2^2)\,q_2+c_1\Bigl(x_1(J_1^2-J_2^2)+2x_2J_1J_2\Bigr)}\,.
\eq
In fact  we have to substitute $q_2$ instead of $q_1$ only.

So, one gets the canonical transformation  from the initial physical variables $(x,J)$ to the variables of separation $(q,p)$  using a pair of compatible bivectors $P^{(1,2)}$ and the corresponding control matrices $F^{(1,2)}$.

In these separated variables entries of the matrix $S$ of normalized eigenvectors of $F^{(1)}$
depend on the pair of variables $(q_i,p_i)$ and the Hamilton function
\[
S_{i1}=-2H_1-\left(4q_i^2-\frac{c_1^2}{4}\right)p_i^2,\qquad S_{1,2}=1,\qquad i=1,2.
\]
In this case $S$ may be called the {\em generalized} St\"ackel matrix. The separated relations  (\ref{stseprel}) look like
\[
S_{i1}H_1+H_2-\left(H_1^2-\dfrac{(c_1^2-16q_i)^2p_i^4}{64}-a^2(c_1^2-16q_i^2)\right)=0,\qquad i=1,2.
\]
and, therefore, the {\em generalized} St\"ackel potential $U_i$ depends on the Hamilton function too.

So, we can say that the variables of separation $(q_i,p_i)$ lie on the algebraic hyperelliptic curve $\mathcal C$ of genus three defined by
\ben
\mathcal C:\qquad \Phi(q,p)&=&
\left(\frac{(c_1^2-16q^2)p^2}{8}-H_1-\sqrt{H_2}\right)
\left(\frac{(c_1^2-16q^2)p^2}{8}-H_1+\sqrt{H_2}\right)\nn\\
&-&a^2(c_1^2-16q^2)=0\,.\label{curve-1}
\en
This equation is invariant with respect to involution $(q,p)\to(-q,p)$. Factorization
with respect to this involution give rise to elliptic curve
\ben
\mathcal E:\qquad \Phi(z,p)&=&
\left(\frac{(c_1^2-16z)p^2}{8}-H_1-\sqrt{H_2}\right)
\left(\frac{(c_1^2-16z)p^2}{8}-H_1+\sqrt{H_2}\right)\nn\\
&-&a^2(c_1^2-16z)=0\,,\qquad z=q^2\label{curve-e1}
\en
Due to the standard formalism  we have to calculate differential on this curve
\[
\Omega=\dfrac{dz}{Z(z,p)},\qquad Z(z,p)=p\,(c_1^2-16z)\Bigl(8H_1-p^2(c_1^2-16z)\Bigl).
\]
Then it's easy to prove that
\bq\label{lin-1}
\dfrac{\dot{q_1}}{Z(q_1^2,p_1)}+\dfrac{\dot{q_2}}{Z(q_2^2,p_2)}=0\,,
\qquad
\dfrac{(c_1^2-16q_1^2)p_1^2\dot{q_1}}{Z(q_1^2,p_1)}
+\dfrac{(c_1^2-16q_2^2)p_2^2\dot{q_2}}{Z(q_2^2,p_2)}=-\dfrac{1}{4}
\eq
and
\ben
&&\int^{q_1} \dfrac{dq}{Z(q^2,p)}+\int^{q_2} \dfrac{dq}{Z(q^2,p)}=\beta_1\,,\nn\\
\nn\\
&&\int^{q_1} \dfrac{(c_1^2-16q^2)p^2dq}{Z(q^2,p)}+\int^{q_2} \dfrac{(c_1^2-16q^2)p^2dq}{Z(q^2,p)}
=-\dfrac{t}{4}+\beta_2\nn
\en
where $p$ has to be obtained from (\ref{curve-1}).

\begin{rem}
The equations of motion are linearized on an abelian variety, which  is roughly spiking the \textit{complexified} of the corresponding Liouville real torus. So, even though $q_{1,2}$ are the real variables of separation we have to solve the Jacobi inversion problem over the complex field, see more detailed discussion in \cite{au96,dub}.
\end{rem}

\begin{rem}
We have to point out the Kowalevski separation of variables leading to hyperelliptic quadratures,
whereas in the new variables of separation $q_{1,2}$
equations of motion  are integrable by quadratures in terms of elliptic functions.
\end{rem}

The third part of the Jacobi method consists of the construction of new integrable systems starting with known variables of separation and some other  separated relations  \cite{jac66}.
If we substitute our variables of separation $(q,p)$ into the following deformation of  (\ref{curve-1})
\bq\label{def-1}
\Phi^{(d)}(p,q)=\Phi(p,q)-8d_1q-16d_2q^2=0,\qquad d_1,d_2\in\mathbb R,
\eq
we get the following generalization of the initial Hamilton function
\bq\label{hd-kow}
H_1^{(d)}= J_1^2+J_2^2+2J_3^2+c_1x_1^2+\dfrac{d_1}{\sqrt{x_1^2+x_2^2}}+\dfrac{d_2}{x_3^2}\,(J_1^2+J_2^2)\,.
\eq
Here the main problem is how to get the Hamiltonian to be  interesting to physics. For example, in our case we obtained the natural Hamiltonian at $d_2=0$ only. For this system only the integrals of motion have been known \cite{yeh}.

The apparent problem is that generalized equations (\ref{def-1}) have not
involution $(q,p)\to(-q,p)$ at $d_1\neq 0$. Thereby equations of motion are related with  the hyperelliptic curve of genus three \cite{matpr08,nak08} instead
of elliptic curve. Nevertheless, at $d_1\neq0$, $d_2=0$ we have the same equations (\ref{lin-1}) as above
\ben
\dfrac{\dot{q_1}}{Z(q_1^2,p_1)}+\dfrac{\dot{q_2}}{Z(q_2^2,p_2)}&=&0\,,\nn\\
\label{lin-2}\\
\dfrac{(c_1^2-16q_1^2)p_1^2\dot{q_1}}{Z(q_1^2,p_1)}
+\dfrac{(c_1^2-16q_2^2)p_2^2\dot{q_2}}{Z(q_2^2,p_2)}&=&-\dfrac{1}{4}\nn
\en
where $p_{1,2}$ satisfy to the deformed equations (\ref{def-1}).

\subsection{The complex Poisson bivector}
For the cubic in momenta Poisson bivector $P^{(3)}$ (\ref{kow-sol3})  the control matrix is equal to
\[
 F^{(3)}=\left(
           \begin{array}{cc}
            2(J_1^2+J_2^2+2J_3^2)+ c_1(x_1+\ii x_2) & -\dfrac{1}{2} \\
            2(J_1^2+J_2^2)^2-2c_1(x_1-\ii x_2)(J_1+\ii J_2)^2 & 0
           \end{array}
         \right)
\]
and the Darboux-Nijenhuis coordinates $\lambda_{1,2}$ are the roots of the characteristic polynomial
\bq\label{q3-def}
\det(F-\lambda I)=(\lambda-\lambda_1)(\lambda-\lambda_2)=\lambda^2-F^{(3)}_{11}\lambda+\dfrac{F^{(3)}_{21}}{2}
\,.
\eq
As above we can get the conjugated  momenta $\mu_{1,2}$ by using compatible with $P^{(3)}$ bivector of fourth order in momenta $J_k$. However we can do it without such calculations as well.

It is easy to see that in this case matrix $S$ of normalized eigenvectors of $F^{(3)}$
is the standard St\"ackel matrix
\[
S_{i1}=-2\lambda_i,\qquad S_{i,2}=1,\qquad i=1,2,
\]
and, therefore, the St\"ackel potentials
\[U_{1,2}=-(S_{i1}H_1+H_2)
\]
are some functions on $(\lambda_1,\mu_1)$ and $(\lambda_2,\mu_2)$, respectively.

In fact notion of the St\"ackel potentials allows us to find the unknown conjugated momenta $\mu_{1,2}$ using the Poisson brackets only. Namely, the following recurrence chain of the Poisson brackets
\bq\label{rrel2}
\phi_1=\{\lambda_1,U_1\},\qquad \phi_2=\{\lambda_1,\phi_1\},\ldots,\quad \phi_i=\{\lambda_1,\phi_{i-1}\}
\eq
is a quasi-periodic chain
\[\phi_3=16 \lambda_1 \phi_1.\]
It means that the St\"ackel potential $U_1$ is a trigonometric function on momenta $\mu_1$ and, therefore, we can determine this desired momenta
\[
\mu_1=\varphi(\lambda_1)\ln\Bigl(\sqrt{16 \lambda_1}\,\phi_1+\phi_2\Bigr)
\]
up to canonical transformations $\mu_1\to \mu_1+\mathrm{g}(\lambda_1)$. Here the function $\varphi(\lambda_1)$ is easily calculated from $\{\lambda_1,\mu_1\}=1$.

These variables of separation $(\lambda_i,\mu_i)$ lie on the hyperelliptic curve of genus three
\bq\label{curve2}
\widetilde{\mathcal C}:\qquad \widetilde{\Phi}(\lambda,\mu)=
\ee^{4\ii \sqrt{\lambda}\mu}+\dfrac{a^4c_1^4}{16}\,\ee^{-4\ii \sqrt{\lambda}\mu}
+\lambda^2-2H_1\lambda+H_2=0\,.
\eq
According to \cite{mark01}, this curve $\widetilde{\mathcal C}$ are related with an elliptic curve $\widetilde{\mathcal E}$ and equations of motion for the Kowalevski top are
linearized on the corresponding abelian variety.

One main difference is that the variables of separation $\lambda_{1,2}$ are complex functions on the initial variables $(x,J)$, whereas $q_{1,2}$ are real functions on them.  It will be important when we express initial real variables via real or complex variables of separation after solving of the Jacobi inversion problem over the complex field.

The other difference is that the affine relations of separations (\ref{curve2}) allows us to study quantum counterpart of the Kowalevski top  \cite{kuzts89,sm00}. For  real variables of separation the procedure of quantization is unknown.

\begin{rem}
In framework of the Sklyanin formalism \cite{Skl} variables of separation are the poles of the Baker-Akhiezer  function with  suitable normalization. In \cite{kuzts89,ts02} we find  such variables of separation $u_{1,2}$  for the Ko\-wa\-lev\-ski-Goryachev-Chaplygin gyrostat
\bq\label{g-Ham}
\widehat{H}_1=J_1^2+J_2^2+2J_3^2+\rho J_3+c_1 x_1+c_2(x_1^2-x_2^2)+c_3x_1x_2+\dfrac{c_4}{x_3^2}
\eq
using $2\times 2$ Lax matrix, its Baker-Akhiezer vector-function and the reflection equation algebra.

It is easy to prove that the Darboux-Nijenhuis variables $\lambda_{1,2}$ (\ref{q3-def}) are related with the  poles $u_{1,2}$ of the Baker-Akhiezer  function by the following point transformation
\bq\label{c-var}
\lambda_{1,2}=u_{1,2}^2\,,
\eq
which gives rise to a ramified two-sheeted covering of $\widetilde{\mathcal C}$, see \cite{mark01}.
\end{rem}

\section{Conclusion}
Starting with the integrals of motion for the Kowalevski top  we found three polynomial in momenta Poisson bivectors, which are compatible with the canonical Poisson bivector on the cotangent bundle $T^*\mathcal S^2$ of two-dimensional sphere.

Then  in framework of the bi-hamiltonian geometry we get new real variables of separation $(q,p)$ for the Kowalevski top on the sphere and reproduce known complex variables $(\lambda,\mu)$. These variables are  related by the canonical transformation
\[
\lambda_{1,2}=\lambda_{1,2}(q_1,q_2,p_1,p_2), \qquad \mu_{1,2}=\mu_{1,2}(q_1,q_2,p_1,p_2),
\]
which may be  rewritten  as  a quasi-point canonical transformation \cite{ts96}
\[
\lambda_{1,2}=\lambda_{1,2}(q_1,q_2,H_1,H_2),
\]
which relates  two hyperelliptic curves of genus three.
We can assume that it is no rational cover  and that these Jacobians  are non-isogeneous in Richelot sense \cite{rich36}.  Similar transformations relate these curves with the Kowalevski curve of genus two. Further inquiry of such relations between hyperelliptic curves goes beyond the scope of this paper, see discussion in \cite{au96,curv3,mark01}.

The proposed approach may be useful for the investigation of  other integrable systems with integrals of motion higher order in momenta, for instance, the search of another real variables of separation for the Ko\-wa\-lev\-ski-Goryachev-Chaplygin gyrostat and its various generalizations \cite{yeh}.

\section{Acknowledgement}
We would like to thank Yu.N. Fedorov and A.V. Bolsinov for helpful  discussions.

\end{document}